\documentclass[12pt]{article}
\usepackage{amsfonts}

\usepackage{graphicx}
\usepackage{amsmath}
\usepackage{float}

\textwidth=6.5in \hoffset=-.75in \voffset=-1in \numberwithin{equation}{section}
\numberwithin{figure}{section}

\input{tcilatex}

\begin{document}

\begin{titlepage}
\bigskip \begin{flushright}
hep-th/0405148\\
\end{flushright}
\vspace{1cm}
\begin{center}
{\Large \bf {New Reducible Five-brane Solutions in M-theory}}\\
\end{center}
\vspace{2cm}
\begin{center}
R. Clarkson{ \footnote{ EMail: r2clarks@sciborg.uwaterloo.ca}}, A.M.
Ghezelbash{ \footnote{ EMail: amasoud@avatar.uwaterloo.ca}}, R. B. Mann{
\footnote{ EMail: mann@avatar.uwaterloo.ca}}\\
Department of Physics, University of Waterloo, \\
Waterloo, Ontario N2L 3G1, Canada\\
\vspace{1cm}
\end{center}

\begin{abstract}
We construct new M-theory solutions of M5 branes that are a realization
of  the fully  localized ten dimensional NS5/D6 and NS5/D5 brane intersections.
These solutions are obtained by embedding self-dual geometries lifted to
M-theory. We reduce these solutions down to ten dimensions, obtaining new
D-brane systems in type IIA/IIB supergravity. The worldvolume theories of the
NS5-branes are new non-local, non-gravitational, six dimensional, T-dual little
string theories with eight supersymmetries.
\end{abstract}
\end{titlepage}\onecolumn   
\bigskip

\section{Introduction}

Fundamental M-theory in the low-energy limit is generally believed to be
effectively described by $D=11$ supergravity \cite{gr1}. This suggests that
brane solutions in the latter theory furnish classical soliton states of
M-theory, motivating considerable interest in this subject. There is
particular interest in supersymmetric $p$-brane solutions that saturate the
BPS bound upon reduction to 10 dimensions.

Recently interesting new supergravity solutions for localized D2/D6 and
D2/D4 intersecting brane systems were obtained \cite{hashi,CGMM2}. By
lifting a D6-(D4-)brane to four-dimensional Taub-NUT/Bolt and Eguchi-Hanson
geometries embedded in M-theory, these solutions were constructed by placing
M2-branes in the Taub-NUT/Bolt and Eguchi-Hanson background geometries. The
special feature of these constructions\ is that the solution is not
restricted to be in the near core region of the D6-(D4-)brane.

\ Since the building blocks of M-theory are M2- and M5-branes, it is natural
to investigate the possibility that this type of construction could be
extended to the 5-brane case. \ This is the subject of the present paper, in
which we consider the embedding of \ Taub-NUT/Bolt and Eguchi-Hanson
geometries in M-theory with a single M5-brane. We find three different
solutions of interest, two of which preserve 1/4 of the supersymmetry. We
then compactify these solutions on a circle, obtaining the different fields
of type IIA string theory. Explicit calculation shows that in all cases the
metric is asymptotically (locally) flat, though for some of our compactified
solutions the type IIA dilaton field diverges at infinity.

The outline of our paper is as follows. In section \ref{sec:review}, we
discuss briefly the field equations of supergravity, the M5-brane metric and
the Killing spinor equations. In section \ref{sec:susy}, we present the
solutions that preserve some of the supersymmetry. We begin by considering
four-dimensional Taub-NUT space (in section \ref{sec:susyTN}), for which we
find two NS5/D6 intersecting solutions. Another metric with similar
self-dual properties is the Eguchi-Hanson (EH) metric, which we consider in
section \ref{sec:susyEH}; we obtain two more NS5/D6 intersecting brane
solutions for this case. In section \ref{sec:TDUAL}, we apply T-duality
transformations on type IIA solutions and find type IIB NS5/D5 intersecting
brane solutions. In section \ref{sec:nonsusy}, we present solutions that do
not preserve any supersymmetry. We discuss the embedding of four-dimensional
Taub-Bolt space in M-theory and show that the partial differential equation
can again be separated into two ordinary differential equations. The
solution of one of these differs from the Taub-NUT case; although an
analytic solution is not possible, we find a numerical solution of that
equation. In section \ref{sec:declim}, we consider the decoupling limit of
our solutions and find evidence that in the limit of vanishing string
coupling, the theory on the worldvolume of the NS5-branes is a new little
string theory.

\section{Review of M5-branes and KK Reduction}

\label{sec:review}

The general Lagrangian and the equations of motion for eleven dimensional
supergravity are given in \cite{DuffKK}. The ones relevant for us here are
when we have maximal symmetry (i.e., for which the expectation values of the
fermion fields is zero); these are%
\begin{eqnarray}
R_{mn}-\frac{1}{2}g_{mn}R &=&\frac{1}{3}\left[ F_{mnpq}F_{n}^{\phantom{n}%
pqr}-\frac{1}{8}g_{mn}F_{pqrs}F^{pqrs}\right]  \label{GminGG} \\
\nabla _{m}F^{mnpq} &=&-\frac{1}{576}\varepsilon ^{m_{1}\ldots
m_{8}npq}F_{m_{1}\ldots m_{4}}F_{m_{5}\ldots m_{8}}  \label{dF}
\end{eqnarray}%
where in what follows $m,n,\ldots $ are 11-dimensional world space indices,
and $a,b,\ldots $ are 11-dimensional tangent space indices. For our purposes
here, with a general M5-brane, the metric and four-form field strength can
be written (see also, for example, \cite{gauntlett}) 
\begin{eqnarray}
ds^{2} &=&H(y,r)^{-1/3}\left( -dt^{2}+dx_{1}^{2}+\ldots +dx_{5}^{2}\right)
+H(y,r)^{2/3}\left( dy^{2}+ds_{4}^{2}(r)\right) ~~~  \label{ds11general} \\
F_{m_{1}\ldots m_{4}} &=&\frac{\alpha }{2}\epsilon _{m_{1}\ldots
m_{5}}\partial ^{m_{5}}H  \label{Fgeneral}
\end{eqnarray}%
where $ds_{4}^{2}(r)$ is a four-dimensional (Euclideanized) metric written
in spherical coordinates, depending on the radius $r$ and the quantity $%
\alpha =\pm 1,$\ which corresponds to an M5-brane and an anti-M5-brane
respectively. The general solution, where the transverse coordinates are
given by a flat metric, admits a solution with 16 Killing spinors \cite%
{gauntlett}. In the sequel we demonstrate that, though the number of Killing
spinors is reduced, we can embed Taub-NUT and Eguchi-Hanson metrics into
this eleven dimensional M5-brane formula to give supersymmetric solutions.

By embedding a Taub-NUT or Eguchi-Hanson metric, one achieves a form that
can be easily reduced down to ten dimensions using the following equations 
\begin{eqnarray}
\hat{g}_{mn} &=&\left[ 
\begin{array}{cc}
e^{-2\Phi /3}\left( g_{\alpha \beta }+e^{2\Phi }C_{\alpha }C_{\beta }\right)
& \nu e^{4\Phi /3}C_{\alpha } \\ 
\nu e^{4\Phi /3}C_{\beta } & \nu ^{2}e^{4\Phi /3}%
\end{array}%
\right] \\
\hat{F}_{(4)} &=&\mathcal{F}_{(4)}+\mathcal{H}_{(3)}\wedge dx_{10}.
\label{FKKreduced}
\end{eqnarray}%
Here $\nu $ is the winding number (the number of times the M5-brane wraps
around the compactified dimension) and $x_{10}$ is the eleventh dimension,
on which we compactify. We use hats in the above to differentiate the
eleven-dimensional fields from the ten-dimensional ones that arise from
compactification. $\mathcal{F}_{(4)}$ and $\mathcal{H}_{(3)}$ are the RR
four-form and the NSNS three-form field strengths corresponding to $%
A_{\alpha \beta \gamma }$ and $B_{\alpha \beta }$.

The number of non-trivial solutions to the Killing spinor equation 
\begin{equation}
\partial _{m}\epsilon +\frac{1}{4}\omega _{abm}\Gamma ^{ab}\epsilon +\frac{1%
}{144}\Gamma _{m}^{\phantom{m}npqr}F_{npqr}\epsilon -\frac{1}{18}\Gamma
^{pqr}F_{mpqr}\epsilon =0  \label{killingspinoreq}
\end{equation}%
determine the amount of supersymmetry of the solution, where the $\omega $'s
are the spin connection coefficients, and $\Gamma ^{a_{1}\ldots
a_{n}}=\Gamma ^{\lbrack a_{1}}\ldots \Gamma ^{a_{n}]}$. The $\Gamma ^{a}$
matrices are the eleven dimensional equivalents of the four dimensional
Dirac gamma matrices, and must satisfy the Clifford algebra 
\begin{equation}
\left\{ \Gamma ^{a},\Gamma ^{b}\right\} =-2\eta ^{ab}.  \label{cliffalg}
\end{equation}

In ten dimensional type IIA string theory, we can have D-branes or
NS-branes. D$p$-branes can carry either electric or magnetic charge with
respect to the RR fields; the metric takes the form \cite{gauntlett}%
\begin{equation}
ds_{10}^{2}=f^{-1/2}\left( -dt^{2}+dx_{1}^{2}+\ldots +dx_{p}^{2}\right)
+f^{1/2}\left( dx_{p+1}^{2}+\ldots +dx_{9}^{2}\right)  \label{gDpbrane}
\end{equation}%
where the harmonic function\thinspace\ $f$ generally depends on the
transverse coordinates.

An NS5-brane carries a magnetic two-form charge; the corresponding metric
has the form 
\begin{equation}
ds_{10}^{2}=-dt^{2}+dx_{1}^{2}+\ldots +dx_{5}^{2}+f\left( dx_{6}^{2}+\ldots
+dx_{9}^{2}\right) .  \label{gNS5brane}
\end{equation}%
In what follows we will obtain a mixture of D-branes and NS-branes.

\section{Supersymmetric Solutions}

\label{sec:susy}

\subsection{Embedding Taub-NUT}

\label{sec:susyTN}

The eleven dimensional M5-brane metric with an embedded Taub-NUT metric has
the following form 
\begin{eqnarray}
ds_{11}^{2} &=&H(y,r)^{-1/3}\left(
-dt^{2}+dx_{1}^{2}+dx_{2}^{2}+dx_{3}^{2}+dx_{4}^{2}+dx_{5}^{2}\right) + 
\notag \\
&&+H(y,r)^{2/3}\left( dy^{2}+ds_{TN_{4}}^{2}\right)  \label{tng11} \\
F_{\psi \theta \phi y} &=&2nr^{2}\sin (\theta )\frac{\partial H}{\partial r}%
~\ \ ~~,~~~~~F_{\psi \theta \phi r}=-2nr(r+2n)\sin (\theta )\frac{\partial H%
}{\partial y}  \label{tnF2}
\end{eqnarray}%
where the Taub-NUT metric $ds_{TN_{4}}^{2}$ is given by 
\begin{eqnarray}
ds_{TN_{4}}^{2} &=&\frac{1}{f(r)}\left( d\psi ^{\prime }+2n\cos (\theta
)d\phi \right) ^{2}+f(r)dr^{2}+(r^{2}-n^{2})\left( d\theta ^{2}+\sin
^{2}(\theta )d\phi ^{2}\right) ~~  \label{tng1} \\
f(r) &=&\frac{(r+n)}{(r-n)}  \label{tnfr}
\end{eqnarray}%
and we choose $\alpha =1.$ An equivalent form of this metric that is more
useful for reduction down to ten dimensions is given by replacing $\psi
^{\prime }=4n\psi $ and $r\rightarrow r+n$, to give 
\begin{eqnarray}
ds_{TN_{4}}^{2} &=&\frac{(4n)^{2}}{\tilde{f}(r)}\left( d\psi +\frac{1}{2}%
\cos (\theta )d\phi \right) ^{2}+\tilde{f}(r)\left( dr^{2}+r^{2}\left(
d\theta ^{2}+\sin ^{2}(\theta )d\phi ^{2}\right) \right)  \label{tng2} \\
\tilde{f}(r) &=&\left( 1+\frac{2n}{r}\right)
\end{eqnarray}%
a form previously utilized for the M2-brane \cite{hashi,CGMM2}, with $\psi $
and $\phi $ having period $2\pi $. The form (\ref{tng1}) is more convenient
for determining how much supersymmetry is preserved, whereas the form (\ref%
{tng2}) is more convenient for solving for $H(y,r)$ and reducing to ten
dimensions.

This metric (\ref{tng11}) is a solution to the eleven dimensional
supergravity equations provided $H\left( y,r\right) $ is a solution to the
differential equation 
\begin{equation}
\frac{r}{2(r+2n)}\frac{\partial ^{2}H}{\partial r^{2}}+\frac{1}{r+2n}\frac{%
\partial H}{\partial r}+\frac{1}{2}\frac{\partial ^{2}H}{\partial y^{2}}=0.
\label{tndiffeq}
\end{equation}%
This equation is straightforwardly separable. Substituting 
\begin{equation}
H(y,r)=1+Q_{M5}Y(y)R(r)  \label{Hyrsep}
\end{equation}%
where $Q_{M5}$ is the charge on the M5-brane, we arrive at two differential
equations 
\begin{eqnarray}
\frac{d^{2}Y(y)}{dy^{2}}+c^{2}Y(y) &=&0  \label{tndeqY} \\
r\frac{d^{2}R(r)}{dr^{2}}+2\frac{dR(r)}{dr}-c^{2}(r+2n)R(r) &=&0.
\label{tndeqR}
\end{eqnarray}%
These have the solutions 
\begin{eqnarray}
Y(y) &=&C_{1}\cos (cy)+C_{2}\sin (cy)  \label{tnYs} \\
R(r) &=&D_{1}e^{-cr}\mathcal{G}(1+cn,2,2cr)+D_{2}e^{-cr}\mathcal{U}%
(1+cn,2,2cr)  \label{tnRs}
\end{eqnarray}%
where $\mathcal{G}$ is the hypergeometric function, and $\mathcal{U}$ is the
Kummer U function \cite{HandbookMathFns}. Requiring the solution to decay
for large $r$ implies $D_{1}=0$, in which case this solution is now exactly
the same as that obtained in the M2 case \cite{hashi,CGMM2}.

For small $r$ the Taub-NUT metric becomes 
\begin{equation}
ds_{TN_{4}}^{2}=dz^{2}+z^{2}d\Omega _{3}^{2}.
\end{equation}%
In the $n\rightarrow \infty $ limit, keeping $z^{2}=8nr$ fixed (i.e. $r\ll n$%
), the function $R(r)$ must become%
\begin{equation}
R(r)\rightarrow D_{2}\frac{2K_{1}(2c\sqrt{2nr})}{c\Gamma (cn)\sqrt{2nr}}.
\label{limitR}
\end{equation}%
The final solution will be a superposition of all possible solutions 
\begin{equation}
H_{TN_{4}}(y,r)=1+Q_{M5}\int dc\{f_{1}(c)\cos (cy)+f_{2}(c)\sin (cy)\}e^{-cr}%
\mathcal{U}(1+cn,2,2cr)  \label{HTN}
\end{equation}%
where $f_{1}(c)$ and $f_{2}(c)$ are functions of $c$ that also contain the
integration constants $C_{i}$, $D_{2}$ from above. In the small $r$ limit,
we require that the above solution satisfies 
\begin{eqnarray}
&&1+Q_{M5}\lim_{z^{2}<<8n^{2}}\int dc\{f_{1}(c)\cos (cy)+f_{2}(c)\sin
(cy)\}e^{-cr}\mathcal{U}(1+cn,2,2cr)  \notag \\
&=&1+Q_{M5}\int dc\{f_{1}(c)\cos (cy)+f_{2}(c)\sin (cy)\}\frac{2K_{1}(2c%
\sqrt{2nr})}{c\Gamma (cn)\sqrt{2nr}}  \notag \\
&=&1+\frac{Q_{M5}}{R^{3}}
\end{eqnarray}%
where $R=\sqrt{y^{2}+z^{2}}=\sqrt{y^{2}+8nr}$. This implies $C_{2}=0$ in (%
\ref{tnYs}) and gives $f_{1}(c)={\frac{c^{2}\Gamma (cn)}{2\pi }}$. The final
solution is 
\begin{equation}
H_{TN_{4}}(y,r)=1+Q_{M5}\int dc\frac{c^{2}\Gamma (cn)}{2\pi }\cos (cy)e^{-cr}%
\mathcal{U}(1+cn,2,2cr).  \label{HfinTN4}
\end{equation}

By reversing the sign of the separation constant $c^{2}$ in equations (\ref%
{tndeqY}) and (\ref{tndeqR}) (so that $c$ $\rightarrow i\widetilde{c}$ ) we
easily obtain another solution . In this case the equation (\ref{tndeqY})
has the decaying solution 
\begin{equation}
Y_{\widetilde{c}}(y)=\widetilde{C}_{\widetilde{c}}e^{-\widetilde{c}y}
\label{SWOlDE2TN4sc}
\end{equation}%
and the radial equation has the solution,

\begin{equation}
R_{\widetilde{c}}(r)=\widetilde{D}_{\widetilde{c}}\frac{(-i)\mathcal{W}%
_{M}(-i\widetilde{c}n,1/2,2i\widetilde{c}r)}{r}=\widetilde{E}_{\widetilde{c}%
}e^{-i\widetilde{c}r}\text{\ }\mathcal{G}(1+i\widetilde{c}n,2,2i\widetilde{c}%
r)  \label{SolDETN4sc}
\end{equation}%
where $\mathcal{G}$\ is a hypergeometric function that is finite at $r=0$,
and undergoes damped oscillations until it vanishes at $r=\infty $. Here $ 
\widetilde{D}_{\widetilde{c}}$ or $\widetilde{E}_{\widetilde{c}}$ are
constants that depending only on $n$ and the separation constant $\widetilde{%
c}$. Writing the general solution of the metric function as a superposition
of solutions yields 
\begin{equation}
\widetilde{H}_{TN_{4}}(y,r)=1+Q_{M5}\int_{0}^{\infty }d\widetilde{c}R_{%
\widetilde{c}}(r)Y_{\widetilde{c}}(y)=1+Q_{M5}\int_{0}^{\infty }d\widetilde{c%
}\widetilde{f}(\widetilde{c})e^{-i\widetilde{c}r}\mathcal{G}(1+i\widetilde{c}%
n,2,2i\widetilde{c}r)e^{-\widetilde{c}y}  \label{HM2secondcase}
\end{equation}%
where $\widetilde{f}(\widetilde{c})=\widetilde{C}_{\widetilde{c}}\widetilde{%
E }_{\widetilde{c}}$. The function $\widetilde{f}(\widetilde{c})$ may again
be determined from a consideration of the near horizon limit where $r<<n$.
This gives%
\begin{eqnarray}
\lim_{z^{2}<<8n^{2}}\int_{0}^{\infty }d\widetilde{c}\widetilde{f}(\widetilde{%
c})e^{-i\widetilde{c}r}\mathcal{G}(1+i\widetilde{c}n,2,2i\widetilde{c}r)e^{-%
\widetilde{c}y} &=&\int_{0}^{\infty }d\widetilde{c}\widetilde{f}(\widetilde{c%
})\frac{I_{1}(2i\widetilde{c}\sqrt{2nr})}{i\widetilde{c}\sqrt{2nr}}e^{-%
\widetilde{c}y}  \notag \\
&=&\frac{1}{(y^{2}+8nr)^{3/2}}
\end{eqnarray}%
and yields $\widetilde{f}(\widetilde{c})=\frac{\widetilde{c}^{2}}{2}$ so that

\begin{equation}
\widetilde{H}_{TN_{4}}(y,r)=1+\frac{Q_{M5}}{2}\int_{0}^{\infty }d\widetilde{c%
}\{e^{-i\widetilde{c}r}\mathcal{G}(1+i\widetilde{c}n,2,2i\widetilde{c}r)e^{-%
\widetilde{c}y}\}\widetilde{c}^{2}.  \label{TN4gfun}
\end{equation}

\begin{figure}[tbp]
\centering        
\begin{minipage}[c]{.3\textwidth}
        \centering
        \includegraphics[width=\textwidth]{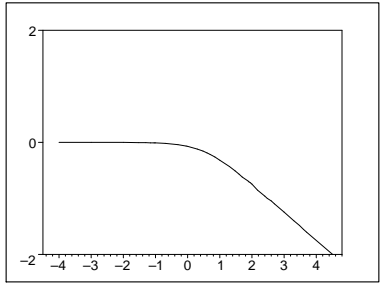}
    \end{minipage}{\large \ }
\caption{Log-Log plot of the functions $h(r)$ in terms of $\frac{r}{n}.$ }
\label{figHTN4}
\end{figure}

Although the terms in the integrand (enclosed by curly brackets) approach a
finite value at $r=y=0$ (unlike the situation for the 2-brane scenario where
the analogous integrand diverges on the brane location \cite{hashi,CGMM2}),
the quantity $\widetilde{h}(y)=\{\widetilde{H}_{TN_{4}}(y,0)-1\}/Q_{M5}=%
\frac{1}{y^{3}}$ diverges at $y=0.$ In figure \ref{figHTN4}, a log-log plot
of $h(r)\simeq (H(y=0,r)-1)$\ versus $\frac{r}{n}$\ is given, where we
choose the normalization coefficient such that it approaches unity as $r$\
goes to zero. A log-log plot of a similarly normalized $\widetilde{h}(y)$
clearly yields a horizontal line. 

We can reduce the metric (\ref{tng11}) down to ten dimensions using the
Kaluza-Klein prescription. The radius $R_{\infty }$ of the circle of
compactification with line element $R_{\infty }^{2}\left[ d\psi +{\textstyle%
\frac{1}{2}}\cos (\theta )d\phi \right] ^{2}$ is 
\begin{equation}
R_{\infty }=4n=g_{s}\ell _{s}  \label{tn4R}
\end{equation}%
using the form in eq. (\ref{tng2}). Reducing to ten dimensions gives the
following NSNS dilaton 
\begin{equation}
\Phi =\frac{3}{4}\ln \left\{ \frac{H^{2/3}}{\nu ^{2}\tilde{f}}\right\} .
\label{dilTN}
\end{equation}%
The NSNS field strength of the two-form associated with the NS5-brane, found
from (\ref{FKKreduced}), is given by 
\begin{equation}
\mathcal{H}_{(3)}=\frac{F_{\theta \phi y\psi }}{4n}d\theta \wedge d\phi
\wedge dy+\frac{F_{\theta \phi r\psi }}{4n}d\theta \wedge d\phi \wedge dr
\label{tnH3}
\end{equation}%
where $F_{\theta \phi y\psi }$ and $F_{\theta \phi r\psi }$ are given in (%
\ref{tnF2}). \ From (\ref{tnH3}), the NSNS two form is given by%
\begin{equation}
B_{(2)}=r^{2}\cos \theta \frac{\partial H}{\partial r}dy\wedge d\phi
+r(r+2n)\cos \theta \frac{\partial H}{\partial y}d\phi \wedge dr.
\label{IIANSNStwoform}
\end{equation}%
The RR fields are 
\begin{eqnarray}
C_{(1)} &=&2n\cos (\theta )d\phi  \label{tnRR} \\
\mathcal{A}_{\alpha \beta \gamma } &=&0  \label{tnA3}
\end{eqnarray}%
where $C_{\alpha }$ is the field associated with the D6-brane, and the
metric in ten dimensions is given by: 
\begin{eqnarray}
ds_{10}^{2} &=&\frac{1}{\nu }\Bigg(\tilde{f}^{-1/2}\left(
-dt^{2}+dx_{1}^{2}+dx_{2}^{2}+dx_{3}^{2}+dx_{4}^{2}+dx_{5}^{2}\right) +H%
\tilde{f}^{-1/2}dy^{2}+  \notag \\
&&+H\tilde{f}^{1/2}\left( dr^{2}+r^{2}d\Omega _{2}^{2}\right) \Bigg).
\label{tng10}
\end{eqnarray}%
From (\ref{tnRR}), (\ref{tnH3}) and the metric, we can see that the above
ten dimensional metric is an NS5$\perp $D6(5) brane solution.{\large \ }We
have explicitly checked that the above 10-dimensional metric (\ref{tng10}),
with the other fields (the dilaton (\ref{dilTN}), the 1-form field (\ref%
{tnRR}), and the NSNS field strength (\ref{tnH3})) furnish a solution to the
10-dimensional supergravity equations of motion. \ \bigskip

As with the 2-brane scenario, the solution (\ref{tng11}) preserves 1/4 of
the supersymmetry. Inserting $\epsilon =H^{-1/12}\epsilon ^{\prime }$ into
eq. (\ref{killingspinoreq}) we find that the presence of the brane implies
the projection 
\begin{equation}
\left( 1-\Gamma ^{\hat{\psi}\hat{\theta}\hat{\phi}\hat{r}\hat{y}}\right)
\epsilon =0  \label{tnproja}
\end{equation}%
or, because we are in eleven dimensions, 
\begin{equation}
\left( 1-\Gamma ^{\hat{t}\hat{x}_{1}\hat{x}_{2}\hat{x}_{3}\hat{x}_{4}\hat{x}%
_{5}}\right) \epsilon =0  \label{tnprojb}
\end{equation}%
and we are left with 1/2 of the supersymmetry. The remaining equations are%
\begin{eqnarray}
\partial _{\psi }\epsilon -\frac{1}{4f^{2}}\frac{df}{dr}\Gamma ^{\hat{\psi}%
\hat{r}}\epsilon +\frac{n}{2(r^{2}-n^{2})f}\Gamma ^{\hat{\theta}\hat{\phi}%
}\epsilon &=&0  \label{tndpsi} \\
\partial _{\theta }\epsilon +\frac{n}{2(r^{2}-n^{2})^{1/2}f^{1/2}}\Gamma ^{%
\hat{\psi}\hat{\phi}}\epsilon -\frac{r}{2(r^{2}-n^{2})^{1/2}f^{1/2}}\Gamma ^{%
\hat{r}\hat{\theta}}\epsilon &=&0  \label{tndtheta} \\
\partial _{\psi }\epsilon -\frac{n\cos (\theta )}{2f^{2}}\frac{df}{dr}\Gamma
^{\hat{\psi}\hat{r}}\epsilon -\frac{n\sin (\theta )}{%
2(r^{2}-n^{2})^{1/2}f^{1/2}}\Gamma ^{\hat{\psi}\hat{\theta}}\epsilon -\frac{%
r\sin (\theta )}{2(r^{2}-n^{2})^{1/2}f^{1/2}}\Gamma ^{\hat{r}\hat{\phi}%
}\epsilon - &&  \notag \\
-\frac{\cos (\theta )}{2}\Gamma ^{\hat{\theta}\hat{\phi}}\epsilon +\frac{%
n^{2}\cos (\theta )}{(r^{2}-n^{2})f}\Gamma ^{\hat{\theta}\hat{\phi}}\epsilon
&=&0  \label{tndphi}
\end{eqnarray}%
and they have the solution%
\begin{equation}
\epsilon =\exp \left\{ -\frac{\theta }{2}\Gamma ^{\hat{\psi}\hat{\phi}%
}\right\} \exp \left\{ \frac{\phi }{2}\Gamma ^{\hat{\theta}\hat{\phi}%
}\right\} \tilde{\epsilon}  \label{tnlorentz}
\end{equation}%
where 
\begin{equation}
\Gamma ^{\hat{\psi}\hat{r}\hat{\theta}\hat{\phi}}\epsilon =+\epsilon
\label{tnproj2}
\end{equation}%
so that 1/4 of the supersymmetry is preserved.

\subsection{Embedding Eguchi-Hanson}

\label{sec:susyEH}

Another metric that lends itself to KK reduction is the Eguchi-Hanson (EH)
metric. We consider the embedding%
\begin{eqnarray}
ds_{11}^{2} &=&H(y,r)^{-1/3}\left( -dt^{2}+dx_{1}^{2}+\ldots
+dx_{5}^{2}\right) +H(y,r)^{2/3}\left( dy^{2}+ds_{EH}^{2}\right)
\label{ehg11} \\
F_{\psi \theta \phi y} &=&\frac{\sin (\theta )(r^{4}-a^{4})}{16r}\frac{%
\partial H}{\partial r}~\ \ ~~,~~~~~F_{\psi \theta \phi r}=-\frac{r^{3}\sin
(\theta )}{16}\frac{\partial H}{\partial y}  \label{ehF}
\end{eqnarray}%
where 
\begin{eqnarray}
ds_{EH}^{2} &=&\frac{r^{2}}{4g(r)}\left[ d\psi +\cos (\theta )d\phi \right]
^{2}+g(r)dr^{2}+\frac{r^{2}}{4}\left( d\theta ^{2}+\sin ^{2}(\theta )d\phi
^{2}\right)  \label{ehg} \\
g(r) &=&\left( 1-\frac{a^{4}}{r^{4}}\right) ^{-1}  \label{ehfr}
\end{eqnarray}%
is the EH metric.

\begin{figure}[tbp]
\centering        
\begin{minipage}[c]{.3\textwidth}
        \centering
        \includegraphics[width=\textwidth]{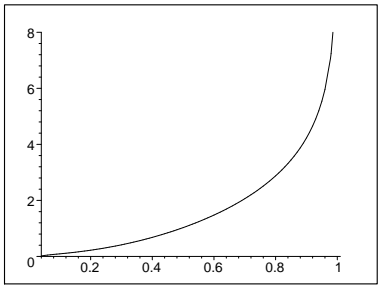}
    \end{minipage}
\caption{ Numerical solution of radial equation (\ref{ehdeqR}) for $R/10^{6}$
in EH case, as a function of $\frac{a}{r}.$ So for $r\approx a$, $R$
diverges and for $r\approx \infty $, it vanishes.}
\label{figEH}
\end{figure}

Eqs. (\ref{ehg11},\ref{ehF}) solve the supergravity equations provided 
\begin{equation}
\frac{(3r^{4}+a^{4})}{2r^{5}}\frac{\partial H}{\partial r}+\frac{%
(r-a)(r+a)(r^{2}+a^{2})}{2r^{4}}\frac{\partial ^{2}H}{\partial r^{2}}+\frac{1%
}{2}\frac{\partial ^{2}H}{\partial y^{2}}=0.  \label{ehdiffeq}
\end{equation}%
This is separable; substituting in (\ref{Hyrsep}) gives two differential
equations 
\begin{eqnarray}
\frac{d^{2}Y(y)}{dy^{2}}+c^{2}Y(y) &=&0  \label{ehdeqY} \\
r(r^{4}-a^{4})\frac{d^{2}R(r)}{dr^{2}}+(3r^{4}+a^{4})\frac{dR(r)}{dr}%
-c^{2}r^{5}R(r) &=&0.  \label{ehdeqR}
\end{eqnarray}%
Eq. (\ref{ehdeqY}) has the solution 
\begin{equation}
Y(y)=C_{1}\cos (cy)+C_{2}\sin (cy)  \label{ehY}
\end{equation}%
and Eq. (\ref{ehdeqR}) doesn't have a solution in terms of known analytic
functions unless $c=0$. However the power-series solutions near $r=a$ are

\begin{eqnarray}
R(r) &=&\left( \tilde{C}_{1}\ln \left( \frac{r}{a}-1\right) \right) \left[ 1+%
\frac{c^{2}a^{2}}{4}\left( \frac{r}{a}-1\right) +\frac{c^{2}a^{2}}{8}\left( 
\frac{r}{a}-1\right) ^{2}+\frac{c^{4}a^{4}}{64}\left( \frac{r}{a}-1\right)
^{2}+\cdots \right] +  \notag \\
&&+\tilde{C}_{2}\left[ -\frac{1+c^{2}a^{2}}{2}\left( \frac{r}{a}-1\right) -%
\frac{1+c^{2}a^{2}}{8}\left( \frac{r}{a}-1\right) ^{2}-\frac{3c^{4}a^{4}}{64}%
\left( \frac{r}{a}-1\right) ^{2}+\cdots \right] +  \notag \\
&&+\mathcal{O}((r-a)^{3})  \label{series}
\end{eqnarray}%
and the solution of interest logarithmically diverges at $r=a$. A typical
numerical solution of (\ref{ehdeqR}) versus $\frac{a}{r}$\ is given in
figure \ref{figEH}. \ 

For the most general solution, we obtain 
\begin{equation}
H_{EH}(y,r)=1+Q_{M5}\int_{0}^{\infty }dc\{g_{1}(c)\cos (cy)+g_{2}(c)\sin
(cy)\}R_{c}(r).  \label{HEH4}
\end{equation}%
To fix the measure functions $g_{1}(c)$\ and $g_{2}(c)$, we compare the
above relation to that of a metric function of a brane in a five-dimensional
flat metric\ $\mathbb{R}\otimes \mathbb{R}^{2}\otimes S^{2},$\ obtained by
looking at the near horizon limit. We note that for $r=a(1+\epsilon ^{2})$,
where $\epsilon <<1,$\ the metric (\ref{ehg}) reduces to 
\begin{eqnarray}
ds_{r=a(1+\epsilon ^{2})}^{2} &=&a^{2}\{\epsilon ^{2}\left[ d\psi +\cos
(\theta )d\phi \right] ^{2}+d\epsilon ^{2}\}+\frac{a^{2}}{4}\left( d\theta
^{2}+\sin ^{2}(\theta )d\phi ^{2}\right)  \label{dseh4reqa} \\
&\approx &z^{2}d\psi ^{2}+dz^{2}+\frac{a^{2}}{4}\left( d\theta ^{2}+\sin
^{2}(\theta )d\phi ^{2}\right)
\end{eqnarray}%
which is $\mathbb{R}^{2}\otimes S^{2}$\ with the radial length equal to $%
\sqrt{z^{2}+\frac{a^{2}}{4}}$. If we assume that the parameter $a$\ is
small, the differential equation (\ref{ehdeqR}) reduces to{\large \ }%
\begin{equation}
\ddot{R}_{c}+\frac{3}{\widehat{r}}\dot{R}_{c}-c^{2}R_{c}=0
\label{eh4diffeq2Rlimit}
\end{equation}%
where $\widehat{r}=r-a=a\epsilon ^{2}\approx 0$\ and the overdot denotes $%
\frac{d}{d\widehat{r}}$. This equation has the solution

\begin{equation}
R_{c}(\widehat{r})=\frac{A_{c}}{\widehat{r}}K_{1}(c\widehat{r})+\frac{B_{c}}{%
\widehat{r}}I_{1}(c\widehat{r})  \label{eh4diffeq2Rlimitsolution}
\end{equation}%
which will vanish at infinity provided $B_{c}=0$, though it will diverge at $%
\widehat{r}=0$.

\begin{figure}[tbp]
\centering        
\begin{minipage}[c]{.3\textwidth}
        \centering
        \includegraphics[width=\textwidth]{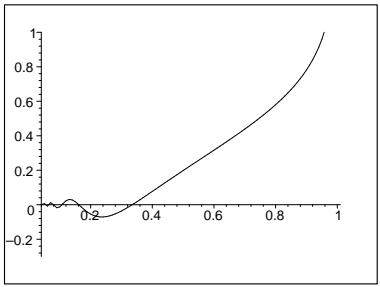}
    \end{minipage}
\caption{Numerical solution of radial equation (\ref{ehdeqR}) with $%
c\rightarrow i\widetilde{c}$ for $\mathcal{R}/(5\times 10^{5})$ in EH case,
as a function of $\frac{a}{r}.$ So for $r\approx a$, $R$ diverges and for $%
r\approx \infty $, it vanishes.}
\label{figEH2}
\end{figure}

Taking these limits into account in equation (\ref{HEH4}), we find that $%
g_{2}(c)=0$ and 
\begin{equation}
\int_{0}^{\infty }dcg_{1}(c)\frac{K_{1}(c\widehat{r})}{\widehat{r}}\cos
(cy)=\lim_{a\rightarrow 0}\frac{1}{(z^{2}+\frac{a^{2}}{4}+y^{2})^{3/2}}=%
\frac{1}{y^{3}}  \label{HEHlim}
\end{equation}%
where we can absorb the constant $A_{c}$\ into the definition of $g_{1}(c)$.
\ By comparing the above relation with the known integral 
\begin{equation}
\int_{0}^{\infty }dcc\frac{K_{1}(c\widehat{r})}{\widehat{r}}\cos (cy)=\frac{%
\pi }{2(\widehat{r}^{2}+y^{2})^{3/2}}\overset{\widehat{r}=a\epsilon ^{2}}{%
\rightarrow }\frac{\pi }{2y^{3}}  \label{K1J1int}
\end{equation}%
we find that $g_{1}(c)=\frac{2c^{2}}{\pi }$, and so we find{\large \ }%
\begin{equation}
H_{EH}(y,r)=1+\frac{Q_{M5}}{\pi }\int_{0}^{\infty }dc\left( 2c^{2}\cos
(cy)R_{c}(r)\right) .  \label{HEH4final}
\end{equation}%
By reversing the sign of the separation constant $c^{2}$ we obtain the other
alternative solution 
\begin{equation}
\widetilde{H}_{EH}(y,r)=1+Q_{M5}\int_{0}^{\infty }\widetilde{f}(\widetilde{c}%
)d\widetilde{c}e^{-\widetilde{c}y}\mathcal{R}_{\widetilde{c}}(r)
\label{HEH4finalsecondsol}
\end{equation}%
where $\mathcal{R}_{\widetilde{c}}(r)$\ is the solution of the equation (\ref%
{ehdeqR}) with $c\rightarrow i\widetilde{c}$ and a typical numerical
solution for $R_{\widetilde{c}}(r)$ versus $\frac{a}{r}$\ is given in figure %
\ref{figEH2}. This solution diverges at $r=a$ and has damped oscillatory
behaviour at large $r$. 
By dimensional analysis, we have $\widetilde{f}(\widetilde{c})=\widetilde{f}%
_{0}\widetilde{c}^{2}.$

The metric (\ref{ehg11}) can also be reduced down to ten dimensions. Doing
so gives the following NSNS dilaton field 
\begin{equation}
\Phi =\frac{3}{4}\ln \left[ \frac{\omega ^{2}H^{2/3}}{4\nu ^{2}g}\right]
\label{dilEH}
\end{equation}%
where we define the dimensionless coordinate $\omega =r/a$ (so that $%
g=g(\omega )$), and field strength (again associated with the 2-form field $%
B_{\alpha \beta }$ of the NS-brane), 
\begin{equation}
\mathcal{H}_{(3)}=\frac{F_{\theta \phi y\psi }}{a}d\theta \wedge d\phi
\wedge dy+\frac{F_{\theta \phi r\psi }}{a}d\theta \wedge d\phi \wedge dr
\label{ehH3}
\end{equation}%
with the coefficients given by (\ref{ehF}). The RR fields are 
\begin{equation}
C_{(1)}=a\cos (\theta )d\phi  \label{ehC1f}
\end{equation}%
\begin{equation}
\mathcal{A}_{\alpha \beta \gamma }=0  \label{ehA3}
\end{equation}%
The metric in ten dimensions is given by 
\begin{eqnarray}
ds_{10}^{2} &=&\frac{\omega }{2\nu }\Bigg\{g^{-1/2}\left(
-dt^{2}+dx_{1}^{2}+dx_{2}^{2}+dx_{3}^{2}+dx_{4}^{2}+dx_{5}^{2}\right) + 
\notag \\
&&+Hg^{-1/2}dy^{2}+Hg^{1/2}a^{2}\left( d\omega ^{2}+\frac{\omega ^{2}}{4g}%
d\Omega _{2}^{2}\right) \Bigg\}  \label{ehg10}
\end{eqnarray}%
which describes an NS5$\perp $D6(5) brane system (D6 because we have a
one-form (\ref{ehC1f}), which is the field for a D6-brane - see \cite%
{Polchinski95})\textbf{\ }that is distinct from the previous case (\ref%
{tng10}). We have explicitly checked that the above 10-dimensional metric (%
\ref{ehg10}), with the other given fields: dilaton (\ref{dilEH}), one form
field (\ref{ehC1f}), and NSNS field strength (\ref{ehH3}), is a solution to
the 10-dimensional supergravity equations of motion.

The metric (\ref{ehg10}) is locally asymptotically flat (though the dilaton
field (\ref{dilEH}) diverges). For large $\omega $\ it\ reduces to 
\begin{equation}
ds_{10}^{2}=\frac{\omega }{2\nu }%
\{-dt^{2}+dx_{1}^{2}+dx_{2}^{2}+dx_{3}^{2}+dx_{4}^{2}+dx_{5}^{2}+dy^{2}+a^{2}(d\omega ^{2}+%
\frac{\omega ^{2}}{4}d\Omega _{2}^{2})\}  \label{ds10Ehlarger}
\end{equation}%
which is a 10D locally flat metric with solid deficit angles. The Kretchmann
invariant of this spacetime vanishes at infinity and is given by 
\begin{equation}
R_{\mu \nu \rho \sigma }R^{\mu \nu \rho \sigma }=\frac{224\nu ^{2}}{\omega
^{6}a^{4}}  \label{KEHlarger}
\end{equation}%
and all the components of the Riemann tensor in the orthonormal basis have
similar $\frac{\nu }{\omega ^{3}a^{2}}$\ behaviour, vanishing at infinity.

In checking the number of supersymmetries in this solution, we again find
the presence of the brane induces the projection (\ref{tnproja}). Insertion
of this relation into (\ref{killingspinoreq}) yields the remaining equations%
\begin{eqnarray}
\partial _{\psi }\epsilon +\frac{1}{4f}\Gamma ^{\hat{\psi}\hat{r}}\epsilon -%
\frac{r}{8f^{2}}\frac{df}{dr}\Gamma ^{\hat{\psi}\hat{r}}\epsilon +\frac{1}{4f%
}\Gamma ^{\hat{\theta}\hat{\phi}}\epsilon &=&0  \label{ehdtheta} \\
\partial _{\theta }\epsilon +\frac{1}{4f^{1/2}}\Gamma ^{\hat{\psi}\hat{\phi}%
}\epsilon -\frac{1}{4f^{1/2}}\Gamma ^{\hat{r}\hat{\theta}}\epsilon &=&0 \\
\partial _{\phi }\epsilon +\frac{\cos (\theta )}{4f}\Gamma ^{\hat{\psi}\hat{r%
}}\epsilon +\frac{\cos (\theta )}{4f}\Gamma ^{\hat{\theta}\hat{\phi}%
}\epsilon -\frac{r\cos (\theta )}{8f^{2}}\frac{df}{dr}\Gamma ^{\hat{\psi}%
\hat{r}}\epsilon - &&  \notag \\
-\frac{\sin (\theta )}{4f^{1/2}}\Gamma ^{\hat{\psi}\hat{\theta}}\epsilon -%
\frac{\sin (\theta )}{4f^{1/2}}\Gamma ^{\hat{r}\hat{\phi}}\epsilon -\frac{%
\cos (\theta )}{2}\Gamma ^{\hat{\theta}\hat{\phi}}\epsilon &=&0
\label{ehdphi}
\end{eqnarray}%
which can easily be solved via%
\begin{equation}
\Gamma ^{\hat{\psi}\hat{r}\hat{\theta}\hat{\phi}}\epsilon =-\epsilon
\label{ehproj2}
\end{equation}%
where 
\begin{equation}
\epsilon =\exp \left\{ \frac{\psi }{2}\Gamma ^{\hat{\psi}\hat{r}}\right\} 
\tilde{\epsilon}.  \label{ehlorentz}
\end{equation}%
Thus, embedding an Eguchi-Hanson metric into the eleven dimensional
equations also gives a 1/4 supersymmetry preserving solution.

\subsection{T-dual IIB configurations}

\bigskip \label{sec:TDUAL}

In this section, we apply T-duality transformations to the IIA NS5$\perp $%
D6(5) configurations. We begin with the case derived from the Taub-NUT
solution, given by fields\textbf{\ }(\ref{dilTN}), (\ref{IIANSNStwoform}), (%
\ref{tnRR}), (\ref{tnA3}) and (\ref{tng10}).

Applying T-duality \cite{Cascales} in the $x_{1}-$direction of the metric (%
\ref{tng10}), gives the IIB dilaton field%
\begin{equation}
\widetilde{\Phi }=\frac{1}{2}\ln \frac{H}{\tilde{f}}  \label{IBdilaton}
\end{equation}%
with the following 10D metric,%
\begin{eqnarray}
\widetilde{ds}_{10}^{2} &=&\tilde{f}^{-1/2}\left( -dt^{2}+\tilde{f}%
dx_{1}^{2}+dx_{2}^{2}+dx_{3}^{2}+dx_{4}^{2}+dx_{5}^{2}\right) +H\tilde{f}%
^{-1/2}dy^{2}+  \notag \\
&&+H\tilde{f}^{1/2}\left( dr^{2}+r^{2}d\Omega _{2}^{2}\right)
\label{IIBmetric}
\end{eqnarray}%
where we set $\nu =1.$\ The only non-zero components of the dual NSNS two
form are given by%
\begin{equation}
\begin{array}{c}
\widetilde{B}_{r\phi }=B_{r\phi } \\ 
\widetilde{B}_{y\phi }=B_{y\phi }%
\end{array}
\label{IIBNSNStwoforms}
\end{equation}%
and so the metric (\ref{IIBmetric}) describes a IIB NS5$\perp $D5(4) brane
configuration.\ We note that the IIB RR axion and four-form fields are zero.
The only non-zero component of the RR two form field turns out to be%
\begin{equation}
\widetilde{\mathcal{B}}_{\phi x_{1}}=C_{\phi }  \label{IIBRRtwoform}
\end{equation}%
The field strength of this two form can be integrated over an 3-sphere. The
result is the charge carried by the D5-brane in a IIB NS5$\perp $D5(4)
system. We have explicitly checked that the IIB metric (\ref{IIBmetric}),
with the NSNS fields (\ref{IBdilaton}), (\ref{IIBNSNStwoforms}), and the RR
two form field (\ref{IIBRRtwoform}) forms a solution to the 10-dimensional
IIB supergravity equations of motion.

A similar situation holds for the IIA NS5$\perp $D6(5) configuration, given
by the metric (\ref{ehg10}) and other fields (\ref{dilEH}), (\ref{ehH3}), (%
\ref{ehC1f}) and (\ref{ehA3}). In this case, T-duality along the $x_{1}-$%
direction of the metric (\ref{ehg10}) with $\nu =1$, gives the IIB dilaton
field%
\begin{equation}
\widetilde{\Phi }=\frac{1}{2}\ln (\frac{\omega ^{2}H}{4g})
\label{IIBdilatonEH}
\end{equation}%
with the following 10D metric,%
\begin{eqnarray}
\widetilde{ds}_{10}^{2} &=&\frac{\omega }{2}\Bigg\{g^{-1/2}\left( -dt^{2}+%
\frac{4g}{\omega ^{2}}dx_{1}^{2}+dx_{2}^{2}+dx_{3}^{2}+dx_{4}^{2}+dx_{5}^{2}%
\right) +  \notag \\
&&+Hg^{-1/2}dy^{2}+Hg^{1/2}a^{2}\left( d\omega ^{2}+\frac{\omega ^{2}}{4g}%
d\Omega _{2}^{2}\right) \Bigg\}  \label{IIBmetricEH}
\end{eqnarray}%
\ The only non-zero components of the dual NSNS two form are given by%
\begin{equation}
\begin{array}{c}
\widetilde{B}_{r\phi }=-\frac{r^{3}\cos \theta }{8a}\frac{\partial H}{%
\partial y} \\ 
\widetilde{B}_{y\phi }=\frac{(r^{4}-a^{4})\cos \theta }{8ra}\frac{\partial H%
}{\partial r}%
\end{array}
\label{IIBNSNStwoformsEH}
\end{equation}%
and so the metric (\ref{IIBmetricEH}) describes a IIB NS5$\perp $D5(4) brane
configuration distinct from the previous case.\ Similar to the preceding
case, the IIB RR axion and four-form fields are zero. The only non-zero
component of the RR two form field turns out to be%
\begin{equation}
\widetilde{\mathcal{B}}_{\phi x_{1}}=a\cos \theta  \label{IIBRRtwoformEH}
\end{equation}%
We have also explicitly checked that the IIB metric (\ref{IIBmetricEH}),
with the NSNS fields (\ref{IIBdilatonEH}) and (\ref{IIBNSNStwoformsEH}), and
the RR two form field (\ref{IIBRRtwoformEH}) provide a solution to the
10-dimensional IIB supergravity equations of motion.

\section{Non-Supersymmetric Solutions}

\label{sec:nonsusy}We consider here embedding the Taub-Bolt version of the
metric (\ref{tng11}). The resultant solutions do not preserve any
supersymmetry. However it has interesting properties that are qualitatively
similar to the previous cases. The metric function $H$ behaves the same way
near the brane core and at infinity, and is an integrated product of a
decaying function and a damped oscillating function far from the brane. Near
the brane core, the convolution of the two functions diverges, as for the
supersymmetric cases.

\subsection{Embedding Taub-Bolt}

\label{sec:nonsusyTB}

The Taub-Bolt metric can be written as%
\begin{eqnarray}
ds_{TB_{4}}^{2} &=&\frac{(4n)^{2}}{\tilde{f}}\left[ d\psi +\frac{\cos
(\theta )}{2}d\phi \right] ^{2}+\tilde{f}dr^{2}+r(r+2n)\left( d\theta
^{2}+\sin ^{2}(\theta )d\phi ^{2}\right)  \label{tbg} \\
\tilde{f}(r) &=&\frac{2r(r+2n)}{(r-n)(2r+n)}  \label{tbfr}
\end{eqnarray}%
and can be embedded into the 11D metric 
\begin{eqnarray}
ds_{11}^{2} &=&H(y,r)^{-1/3}\left(
-dt^{2}+dx_{1}^{2}+dx_{2}^{2}+dx_{3}^{2}+dx_{4}^{2}+dx_{5}^{2}\right) + 
\notag \\
&&+H(y,r)^{2/3}\left( dy^{2}+ds_{TB_{4}}^{2}\right)  \label{TB11d}
\end{eqnarray}%
where the four-form field strength is given by 
\begin{equation}
F_{\psi \theta \phi y}=n\left( 2r^{2}-rn-n^{2}\right) \sin (\theta )\frac{%
\partial H}{\partial r}~~\ \ ~,~~~~~F_{\psi \theta \phi r}=-2(r+2n)rn\sin
(\theta )\frac{\partial H}{\partial y}.  \label{tb4F}
\end{equation}

\begin{figure}[tbp]
\centering        
\begin{minipage}[c]{.3\textwidth}
        \centering
        \includegraphics[width=\textwidth]{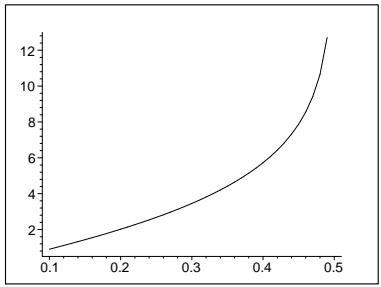}
    \end{minipage}
\caption{ Numerical solution of radial equation (\ref{tbdeqR}) for $R/10^{4}$
in Taub-Bolt$_{4}$ case, as a function of $\frac{n}{2r}.$ So for $r\approx n$%
, $R$ diverges and for $r\approx \infty $, it vanishes.}
\label{figTB}
\end{figure}

An inspection of the metric (\ref{tbg}) indicates that $\psi ,\phi $ each
have period $2\pi $, and that the radius of the circle on which we are
reducing is again $4n$. However, because of the new form (\ref{tbg}), (\ref%
{tbfr}), this is a solution of the eleven dimensional equations of motion
provided that $H(y,r)$ now obeys 
\begin{equation}
\frac{1}{2}\frac{\partial ^{2}H}{\partial y^{2}}+\frac{(4r-n)}{4r(r+2n)}%
\frac{\partial H}{\partial r}+\frac{(r-n)(2r+n)}{4r(r+2n)}\frac{\partial
^{2}H}{\partial r^{2}}=0.  \label{tbdiffeq}
\end{equation}%
This is separable, and using (\ref{Hyrsep}) gives two differential equations
to be solved: 
\begin{eqnarray}
\frac{d^{2}Y(y)}{dy^{2}}+c^{2}Y(y) &=&0  \label{tbdeqY} \\
(r-n)(2r+n)\frac{d^{2}R(r)}{dr^{2}}+(4r-n)\frac{dR(r)}{dr}-2c^{2}r(r+2n)R(r)
&=&0.  \label{tbdeqR}
\end{eqnarray}%
Eq. (\ref{tbdeqY}) has the solution 
\begin{equation}
Y(y)=C_{1}\sin \left( cy\right) +C_{2}\cos \left( cy\right)  \label{tbY}
\end{equation}%
whereas eq. (\ref{tbdeqR}) does not have a solution in terms of known
analytic functions. A power-series solution near the origin yields two
solutions, one of which has a logarithmic divergence at $r=n$. A typical
numerical solution of (\ref{tbdeqR}) versus $\frac{n}{2r}$ is given in
figure \ref{figTB}. 

We note that the most general solution for the metric function is a
superposition of the different solutions of the equations (\ref{tbdeqY}) and
(\ref{tbdeqR}), corresponding to different values of the separation constant 
$c$. We have 
\begin{equation}
H_{TB_{4}}(y,r)=1+Q_{M5}\int_{0}^{\infty }dcp(c)\cos (cy)R_{c}(r)
\label{generalmetricfunctionTB4}
\end{equation}%
where $p(c)=p_{0}c^{2}$ by dimensional analysis, with $p_{0}$\ is a constant
that can be absorbed into the definition of $Q_{M5}$. Since the graph of $%
R_{c}(r)$\ given in figure \ref{figTB} is qualitatively the same as the
corresponding graph for TN$_{4}$, the behaviour of the metric function (\ref%
{generalmetricfunctionTB4}), is similar to the NUT case (\ref{HfinTN4}). The
other solution is%
\begin{equation}
\widetilde{H}_{TB_{4}}(y,r)=1+Q_{M5}\int_{0}^{\infty }d\widetilde{c}~%
\widetilde{p}(\widetilde{c})e^{-\widetilde{c}y}R_{\widetilde{c}}(r)
\label{generalmetricfunctionTB4secondsol}
\end{equation}%
where again $\widetilde{p}(\widetilde{c})=\widetilde{p}_{0}\widetilde{c}%
^{2}. $

Reducing this solution down to ten dimensions gives NSNS and RR fields of
the same form as (\ref{dilTN})-(\ref{tnA3}) (with the $H$ and $\tilde{f}$
now being these new solutions), but the metric in ten dimensions is given by 
\begin{eqnarray}
ds_{10}^{2} &=&\frac{1}{\nu }\Bigg\{\tilde{f}^{-1/2}\left(
-dt^{2}+dx_{1}^{2}+dx_{2}^{2}+dx_{3}^{2}+dx_{4}^{2}+dx_{5}^{2}\right) + 
\notag \\
&&+H\tilde{f}^{-1/2}dy^{2}+H\tilde{f}^{1/2}dr^{2}+H\tilde{f}^{-1/2}r\left(
r+2n\right) d\Omega _{2}^{2}\Bigg\}  \notag \\
&=&\frac{1}{\nu }\Bigg\{\tilde{f}^{-1/2}\left(
-dt^{2}+dx_{1}^{2}+dx_{2}^{2}+dx_{3}^{2}+dx_{4}^{2}+dx_{5}^{2}\right) + 
\notag \\
&&+H\tilde{f}^{-1/2}dy^{2}+H\tilde{f}^{1/2}\left( dr^{\prime 2}+r^{\prime
}\left( r^{\prime }+\frac{3n}{2}\right) d\Omega _{2}^{2}\right) \Bigg\}
\label{tbg10}
\end{eqnarray}%
where we have set $r=r^{\prime }+n$. This describes an NS5$\perp $D6(5)
brane configuration, but again it must be noted that this solution is not
supersymmetric. We have explicitly checked that the above 10-dimensional
metric (\ref{tbg10}), with equations (\ref{dilTN})-(\ref{tnA3}), where $%
\tilde{f}$ is given by (\ref{tbfr}), is a solution to the 10-dimensional
supergravity equations of motion. The metric (\ref{tbg10}) is locally
asymptotically flat; for large $r$\ it\ reduces to 
\begin{equation}
ds_{10}^{2}=\frac{1}{\nu }%
\{-dt^{2}+dx_{1}^{2}+dx_{2}^{2}+dx_{3}^{2}+dx_{4}^{2}+dx_{5}^{2}+dy^{2}+dr^{2}+r(r+2n)d\Omega _{2}^{2})\}
\end{equation}%
which is a 10D locally flat metric. The Kretchmann invariant of this
spacetime vanishes at infinity and is given by 
\begin{equation}
R_{\mu \nu \rho \sigma }R^{\mu \nu \rho \sigma }=\frac{12\nu ^{2}n^{4}}{%
r^{4}(r+2n)^{4}}\sim \frac{1}{r^{8}}
\end{equation}%
and all the components of the Riemann tensor in an orthonormal basis have
similar $\frac{\nu n^{2}}{r^{2}(r+2n)^{2}}\sim \frac{1}{r^{4}}$\ behaviour,
vanishing at infinity.

\section{Decoupling Limits}

\label{sec:declim}

We now wish to perform an analysis of the decoupling limits of the solutions
presented above. As the steps are the same for each case, we will only
present one case (Taub-NUT) in detail, and mention the differences involved
for the other two. Details of the decoupling limit can be found in \cite%
{DecouplingLim}; we also provided a brief review in our previous M2 work %
\cite{CGMM2}.

At low energies, the dynamics of our IIA NS5-branes will decouple from the
bulk. Near the NS5-brane horizon ($H>>1$), we are interested in the
behaviour of the our NS5-branes in the limit where string coupling vanishes 
\begin{equation}
g_{s}\rightarrow 0  \label{gym1}
\end{equation}%
and 
\begin{equation}
\ell _{s}=\text{ fixed.}  \label{gym2}
\end{equation}%
\ In these limits, we rescale the radial coordinates such that they can be
kept fixed 
\begin{equation}
Y=\frac{y}{g_{s}\ell _{s}^{2}}~,~U=\frac{r}{g_{s}\ell _{s}^{2}}.
\label{yrrescale}
\end{equation}%
This will cause the harmonic function of the D6-brane for the Taub-NUT
solutions to change to 
\begin{equation}
\widetilde{f}(r)=1+\frac{2n}{r}=1+\frac{N_{6}}{2U\ell _{s}}\equiv f(U)
\end{equation}%
where we now generalize to $N_{6}$ D6-branes and we have used (\ref{gym2}), (%
\ref{yrrescale}) and (\ref{tn4R}). A similar transformation happens in the
EH and TB cases, where in the EH case the rescaling $a=A\ell _{s}^{2}g_{s}$
is used.

All six of the harmonic functions for the NS5-branes above can be shown to
rescale according to $H(Y,U)=g_{s}^{-2}h(Y,U)$, through the use of (\ref%
{gym2}), (\ref{yrrescale}) (\ref{tn4R}), etc., where appropriate. For
example, the first of our harmonic functions for the Taub-NUT solutions (\ref%
{HfinTN4}) becomes 
\begin{eqnarray}
H_{TN_{4}}(Y,U) &=&Q_{M5}\int dc\frac{c^{2}\Gamma (cn)}{2\pi }\cos
(cy)e^{-cr}\mathcal{U}(1+cn,2,2cr)  \notag \\
&=&\frac{\pi N_{5}}{g_{s}^{2}\ell _{s}^{3}}\int dP~P^{2}\Gamma \left( \frac{P%
}{4\ell _{s}}\right) \cos \left( PY\right) e^{-PU}\mathcal{U}\left( 1+\frac{P%
}{4\ell _{s}},2,2PU\right)  \notag \\
&=&\frac{h(Y,U)}{g_{s}^{2}}
\end{eqnarray}%
where we have used $\ell _{p}=g_{s}^{1/3}\ell _{s}$ to rewrite 
\begin{equation}
Q_{M5}=\pi N_{5}\ell _{p}^{3}=\pi N_{5}g_{s}\ell _{s}^{3}
\end{equation}%
and we have rescaled $c=Pg_{s}^{-1}\ell _{s}^{-2}$ so that $h(Y,U)$ has no $%
g_{s}$ dependence.

The ten-dimensional decoupled metrics are then given by inserting the
appropriate harmonic functions, along with the above limits, into the
metrics. For example, the metric (\ref{tng10}) becomes, in the decoupling
limit 
\begin{eqnarray}
ds_{10}^{2} &=&f^{-1/2}(U)\left(
-dt^{2}+dx_{1}^{2}+dx_{2}^{2}+dx_{3}^{2}+dx_{4}^{2}+dx_{5}^{2}\right) + 
\notag \\
&&+\ell _{s}^{4}\left\{ h(Y,U)f^{-1/2}(U)dY^{2}+h(Y,U)f^{1/2}(U)\left(
dU^{2}+U^{2}d\Omega _{2}^{2}\right) \right\} .  \label{tn10decoupled}
\end{eqnarray}

In the limit of vanishing $g_{s}$\ with fixed $l_{s}$\ (as we did in (\ref%
{gym1}) and (\ref{gym2})), the decoupled free theory on NS5-branes should be
a little string theory \cite{shiraz} (i.e. a 6-dimensional non-gravitational
theory in which modes on the 5-brane interact amongst themselves, decoupled
from the bulk). We note that our NS5/D6 system is obtained from M5-branes by
compactification on a circle of self-dual transverse geometry. Hence the IIA
solution has T-duality with respect to this circle. The little string theory
inherits the same T-duality from IIA string theory, since taking the limit
of vanishing string coupling commutes with T-duality. Moreover T-duality
exists even for toroidally compactified little string theory. In this case,
the duality is given by an $O(d,d,\mathbb{Z})$ symmetry where $d$\ \ is the
dimension of the compactified toroid. These are indications that the little
string theory is non-local at the energy scale $l_{s}^{-1}$ and in
particular in the compactified theory, the energy-momentum tensor can not be
defined uniquely \cite{aha}.

We now consider the analysis of the decoupling limits of the IIB solutions
presented in section \ref{sec:TDUAL}. At low energies, the dynamics of IIB
NS5-branes will decouple from the bulk. Near the NS5-brane horizon ($H>>1$),
the field theory limit is given by 
\begin{equation}
g_{YM5}=\ell _{s}=\text{ fixed}  \label{gym}
\end{equation}%
We rescale the radial coordinates $y\ $and $r$\ as in (\ref{yrrescale}),
such that their corresponding rescaled coordinates $Y$\ and $U$ are kept
fixed. The harmonic function of the D5-brane for the Taub-NUT solutions
becomes 
\begin{equation}
\widetilde{f}(r)=1+\frac{2n}{r}=1+\frac{\pi \widetilde{N}_{5}}{2Ug_{YM5}}%
\equiv \widetilde{f}(U)
\end{equation}%
where $\widetilde{N}_{5}$\ is the number of D5-branes.

The harmonic function of the NS5$\perp $D5 system (\ref{IIBmetric}), given
in eq. (\ref{HfinTN4}) can be shown to rescale according to $%
H(Y,U)=g_{s}^{-2}\widetilde{h}(Y,U)$, where 
\begin{equation}
\widetilde{h}(Y,U)=\frac{N_{5}}{2g_{YM5}^{3}}\int dP~P^{2}\Gamma \left( 
\frac{P}{4g_{YM5}}\right) \cos \left( PY\right) e^{-PU}\mathcal{U}\left( 1+%
\frac{P}{4g_{YM5}},2,2PU\right)
\end{equation}%
In this case, the ten-dimensional metric (\ref{IIBmetric}), in the
decoupling limit becomes%
\begin{eqnarray}
\widetilde{ds}_{10}^{2} &=&\tilde{f}^{-1/2}(U)\left( -dt^{2}+\tilde{f}%
(U)dx_{1}^{2}+dx_{2}^{2}+dx_{3}^{2}+dx_{4}^{2}+dx_{5}^{2}\right)  \notag \\
&&+g_{YM5}^{4}\widetilde{h}(Y,U)\{\tilde{f}^{-1/2}(U)dY^{2}+\tilde{f}%
^{1/2}(U)\left( dU^{2}+U^{2}d\Omega _{2}^{2}\right) \}
\end{eqnarray}

The decoupling limit illustrates that the decoupled theory in the low energy
limit is super Yang-Mills theory with $g_{YM}=\ell _{s}.$\ In the limit of
vanishing $g_{s}$\ with fixed $l_{s}$,\ the decoupled free theory on IIB
NS5-branes (which is equivalent to the limit $g_{s}\rightarrow \infty $\ of
decoupled S-dual of the IIB D5-branes) reduces to a IIB (1,1) little string
theory with eight supersymmetries.\ 

\section{Conclusions}

The central thrust of this paper is the construction of supergravity
solutions for fully localized NS5/D6 brane intersections without restricting
to the near core region of the D6 branes. We have constructed these
solutions by lifting Taub-NUT, Eguchi-Hanson and Taub-Bolt spaces into
M-theory. These exact solutions are new M5 metrics, and are presented in
equations (\ref{HfinTN4}), (\ref{TN4gfun}), (\ref{HEH4final}), (\ref%
{HEH4finalsecondsol}), (\ref{generalmetricfunctionTB4}), and (\ref%
{generalmetricfunctionTB4secondsol}) which are the main results of this
paper.

The common feature of all of these solutions is that the brane function is a
convolution of an exponentially decaying `radial' function with a damped
oscillating one. The `radial' functions vanish far from the M5-brane and
diverge near the brane core. Dimensional reduction to 10 dimensions gives us
different NS5/D6 brane intersections, which in two of our cases - the
Taub-NUT and Eguchi-Hanson spaces (which have self-dual Riemann curvature) -
preserve 1/4 of the supersymmetry and yield metrics with acceptable
asymptotic behaviour. In the last case, involving the 4-dimensional
Taub-Bolt metric, the system is not supersymmetric. However the general
functional structure of the brane is qualitatively the same as for the
supersymmetric cases: the $r$-dependent part of the metric function diverges
for small $r$ and falls off rapidly for large $r$, whereas the $y$-dependent
part of the metric functions is given by an oscillating sine function (or
decaying exponential function). 

We note that all the solutions where the integrands contain $e^{-\widetilde{c%
}y}$, are not convergent for all values of  $y$. To make the integral
convergent for $y<0$, one can replace $e^{-\widetilde{c}y}$\ by $e^{-%
\widetilde{c}\left| y\right| }$, but only at the price of introducing a
source term at $y=0$ in the corresponding Laplace equation for $\widetilde{H}%
(y,r)$.

We considered the decoupling limit of our solutions and found that NS5-brane
can decouple from the bulk. The resulting theory on the NS5-brane in the
limit of vanishing string coupling with fixed string length is a little
string theory.

In the standard case, the system of N$_{5}$\ NS5-branes located at N$_{6}$\
D6-branes can be obtained by dimensional reduction of \ N$_{5}$N$_{6}$\
coinciding images of M5-branes in the flat transverse geometry. In this
case, the worldvolume theory (the little string theory) of the IIA
NS5-branes, in the absence of D6-branes, is a non-local non-gravitational
six dimensional theory \cite{seiberg}. This theory has (2,0) supersymmetry
(four supercharges in the \textbf{4}\ representation of Lorentz symmetry $%
Spin(5,1)$) and an R-symmetry $Spin(4)$ remnant of the original ten
dimensional Lorentz symmetry. The presence of the D6-branes breaks the
supersymmetry down to (1,0), with eight supersymmetries. Since we found that
some of our solutions preserve 1/4 of supersymmetry, we expect that the
theory on NS5-branes is a new little string theory. \ 

By T-dualization of the 10D IIA theory along a direction parallel to the
worldvolume of the IIA NS5, we find a IIB NS5$\perp $D5(4) system,
overlapping in four directions. The worldvolume theory of the IIB
NS5-branes, in the absence of the D5-branes, is a little string theory with
(1,1) supersymmetry. The presence of the D5-brane, which has one transverse
direction relative to NS5 worldvolume, breaks the supersymmetry down to
eight supersymmetries. This is in good agreement with the number of
supersymmetries in 10D IIB theory: T-duality preserves the number of
original IIA supersymmetries, which is eight. Moreover we conclude that the
new IIA and IIB little string theories are T-dual: the actual six
dimensional T-duality is the remnant of the original 10D T-duality after
toroidal compactification.

A useful application of the exact M-brane solutions in our paper is to
employ them as supergravity duals of the NS5 worldvolume theories with
matter coming from the extra branes. More specifically, these solutions can
be used to compute some correlation functions and spectrum of fields of our
new little string theories.

In the standard case of $A_{k-1}$ (2,0) little string theory, there is an
eleven dimensional holographic dual space obtained by taking appropriate
small $g_s$ limit of an M-theory background corresponding to M5-branes with
a transverse circle and $k$ units of 4-form flux on $S^3 \otimes S^1$. In
this case, the supergravity approximation is valid for the (2,0) little
string theories at large $k$ and at energies well below the string scale.
The two point function of the energy-momentum tensor of the little string
theory can be computed from classical action of the supergravity evaluated
on the classical field solutions [10].

Near the boundary of the above mentioned M-theory background, the string
coupling goes to zero and the curvatures are small. Hence it is possible to
compute the spectrum of fields exactly. In reference [11], the full spectrum
of chiral fields in the little string theories was computed and the results
are exactly the same as the spectrum of the chiral fields in the low energy
limit of the little string theories. Moreover, the holographic dual theories
can be used for computation of some of the states in our little string
theories.

We conclude with a few comments about possible directions for future work.
Investigation of the different regions of the metric (\ref{tng11}) or
alternatively the 10D string frame metric (\ref{tn10decoupled}) with a
dilaton (also for other considered EH and TB cases) for small and large
Higgs expectation value $U$\ would be interesting, as it could provide a
means\ for finding a holographical dual relation to the new little string
theory we obtained. Moreover, the Penrose limit of the near-horizon geometry
may be useful for extracting information about the high energy spectrum of
the dual little string theory \cite{Gomis}. The other open issue is the
possibility of the construction of a pp-wave spacetime which interpolates
between the different regions of the our new IIA NS5-branes.

\bigskip

{\Large Acknowledgments}

This work was supported by the Natural Sciences and Engineering Research
Council of Canada.

\end{document}